\documentclass[a4paper,11pt]{article}
\usepackage{pos}
\usepackage{bm,latexsym,amsmath,mathrsfs}
\usepackage{graphicx}
\usepackage{color}
\usepackage{times,setspace}

\title{Chiral fermion on quantum computers}
\author*[a]{Arata~Yamamoto}
\affiliation[a]{Department of Physics, The University of Tokyo, Tokyo 113-0033, Japan}
\author[b,c]{Tomoya Hayata}
\affiliation[b]{Departments of Physics, Keio University, Kanagawa 223-8521, Japan}
\affiliation[c]{RIKEN iTHEMS, RIKEN, Wako 351-0198, Japan}
\author[d]{Katsumasa Nakayama}
\affiliation[d]{RIKEN Center for Computational Science, Kobe 650-0047, Japan}

\abstract{
Quantum computation often suffers from artificial symmetry breaking.
We should strive to suppress the artifact both by theoretical and technological improvements.
The theoretical formalism of the lattice fermion with exact chiral symmetry is called the chiral fermion.
In this presentation, we show how the chiral fermion describes chiral physics in quantum computing.
We also show that, although a drawback of the chiral fermion is large computational cost, there is a loophole in one dimension.
}

\FullConference{The 40th International Symposium on Lattice Field Theory (Lattice 2023)\\
July 31st - August 4th, 2023\\
Fermi National Accelerator Laboratory\\}

\begin{document}
\maketitle

\section{Introduction}

In the quantum computation of lattice gauge theory, symmetry is artificially broken for many reasons; e.g., quantum noise, truncation, lattice discretization, etc.
A famous example is artificial gauge symmetry breaking.
Many artifact-robust schemes, such as gauge fixing and dual variable representation, were suggested.
What about chiral symmetry?
Chiral symmetry is theoretically complicated on a lattice.
The standard lattice fermion, the Wilson fermion, breaks chiral symmetry even on classical computation.
The lattice fermion has to be formulated so as to respect chiral symmetry.
Such a formulation is called the chiral fermion formalism.
Here we discuss the chiral fermion in the Hamiltonian lattice gauge theory and its application to quantum computation.
The contents of this presentation are based on two papers: Sec.~\ref{sec2} is based on Ref.~\cite{Hayata:2023zuk} and Sec.~\ref{sec3} is based on Ref.~\cite{Hayata:2023skf}.

\section{Overlap fermion}
\label{sec2}

The overlap fermion is the most renowned example of the chiral fermion.
The Hamiltonian formalism of the overlap fermion was given long ago \cite{Creutz:2001wp}.
The massless overlap fermion is described by the Hamiltonian
\begin{equation}
 H_f = \psi^\dagger \gamma^0 D \psi
\end{equation}
and the overlap Dirac operator
\begin{equation}
 D = 1 + \frac{D_W}{\sqrt{D_W^\dagger D_W}} 
\end{equation}
with the Wilson Dirac operator $D_W$.
Note that $D$ and $D_W$ are the three-dimensional quantum Dirac operators, not the four-dimensional classical Dirac operators in the path integral formalism.
We can define two types of chiral charge operators: the ``naive'' chiral charge
\begin{equation}
 Q_{\rm naive}= \psi^\dagger \gamma^5 \psi
\end{equation}
and the ``conserved'' chiral charge
\begin{equation}
 Q=\psi^\dagger \gamma^5 \left(1-\frac{1}{2}D\right)\psi .
\end{equation}
We can algebraically show that the naive one does not commute with the Hamiltonian,
\begin{equation}
 [H_f,Q_{\rm naive}]\neq 0 ,
\end{equation}
but the conserved one commutes with the Hamiltonian,
\begin{equation}
\label{eqHQ}
 [H_f,Q]=0 .
\end{equation}
Thus the chiral charge $\langle Q \rangle$ is conserved even for nonzero lattice spacing.
The commutation relation \eqref{eqHQ} or the existence of the conserved chiral charge defines the chiral fermion in the Hamiltonian formalism.
This definition is clear and intuitive.
Unlike in the path integral formalism, we do not need the Ginsparg-Wilson relation to define the chiral fermion.

The commutation relation \eqref{eqHQ} implies that the two operators are simultaneously diagonalizable.
Let us write the eigenvalue equations as
\begin{align}
 H_f|\Psi_n\rangle = \varepsilon_n|\Psi_n\rangle \\
 Q|\Psi_n\rangle = q_n|\Psi_n\rangle .
\end{align}
We can draw the two-dimensional plot, $\varepsilon_n$ vs $q_n$, and analyze how the eigenvalue spectrum reflects the chiral property of a system.
In the continuous theory, the chirality of a Dirac fermion is independent of its energy, and $q_n=\pm 1$ if the fermion is massless.
On the lattice, the energy and the chiral charge are correlated.
The one-particle eigenvalue spectrum satisfies
\begin{equation}
 \left(\frac{\varepsilon_n}{2}\right)^2 + \left(q_n\right)^2 =1
\end{equation}
and thus is distributed on a unit circle.

\begin{figure}[h]
\centering
\includegraphics[scale=0.14]{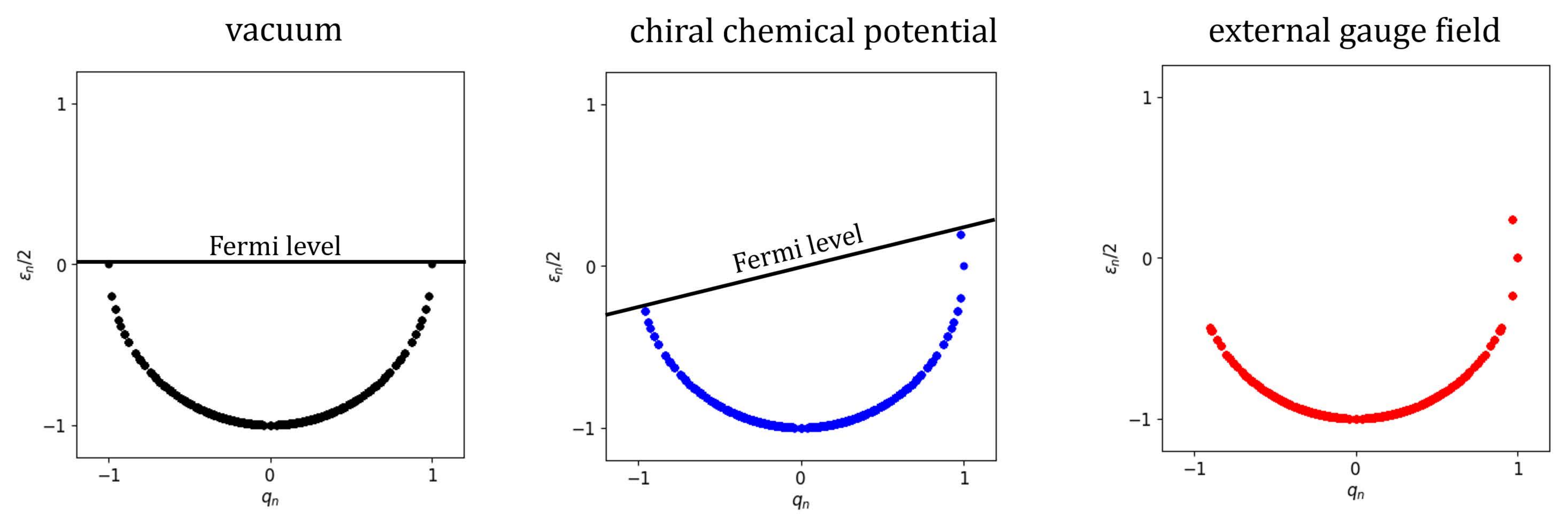}
\caption{\label{fig1}
Eigenvalue spectra of the three-dimensional overlap fermion in the non-interacting vacuum (left), with a chiral chemical potential (center), and with external electromagnetic gauge field (right) \cite{Hayata:2023zuk}.
}
\end{figure}

The physical states can be constructed by filling the one-particle eigenvalue spectrum.
The examples are shown in Fig.~\ref{fig1}.
In the vacuum, the Fermi level is $\varepsilon=0$ and negative energy states are occupied.
The total chiral charge is zero because of parity symmetry, i.e., the inversion symmetry of $q_n$.
We can make the chiral charge nonzero by introducing a chiral chemical potential.
The chiral chemical potential tilts the Fermi level and breaks the inversion symmetry.
A more interesting case is the eigenvalue spectrum with external gauge field.
Let us consider the parallel electric and magnetic fields in the same direction.
(Since the electric field is nonzero, the system evolves in time.
The figure is a snapshot at a certain time.)
The inversion symmetry is broken and the chiral charge is dynamically generated.
This is interpreted as the chiral anomaly in the continuum limit.
This observation is surprising.
The chiral anomaly comes from ultraviolet divergence in the conventional understanding.
On a lattice, however, there is no ultraviolet divergence, or no chiral anomaly.
In the overlap fermion formalism, the chiral anomaly can be observed even on a finite lattice.

In the above analysis, we discussed the non-interacting fermion and the fermion coupled with external (classical) gauge field because the computation is easy.
When the fermions do not interact with each other, we only need the one-particle eigenvalue spectrum of the fermion.
The computational cost is just a polynomial function of the system size.
In general, however, the computation is not so easy.
When the fermion couples with quantum gauge field, we need to treat the full Hilbert space to compute the eigenvalue spectrum.
The computational cost is exponentially large.
The computation gets out of hand on classical computers and we need the help of quantum computers.

\section{Quantum simulation}
\label{sec3}

Let us consider the quantum simulation of real-time evolution
\begin{equation}
\label{eqT}
 |\Psi(t) \rangle = e^{-i(H_g+H_f)t} |\Psi(0) \rangle .
\end{equation} 
In general, the advantage of the overlap fermion is exact chiral symmetry and the disadvantage is large computational cost.
For near-term quantum computers, the simulation of one-dimensional gauge theory is realistic and favored.
In one dimension, there is a special property; the overlap fermion is equivalent to the Wilson fermion \cite{Horvath:1998gq}.
Because of this property, the simulation with exact chiral symmetry is possible.
We just replace the naive chiral charge operator by the conserved chiral charge operator.
We can use the standard quantum circuit of the one-dimensional Wilson fermion for the time evolution \eqref{eqT}.

\begin{figure}[h]
\begin{minipage}{0.5\linewidth}
\centering
\includegraphics[scale=0.75]{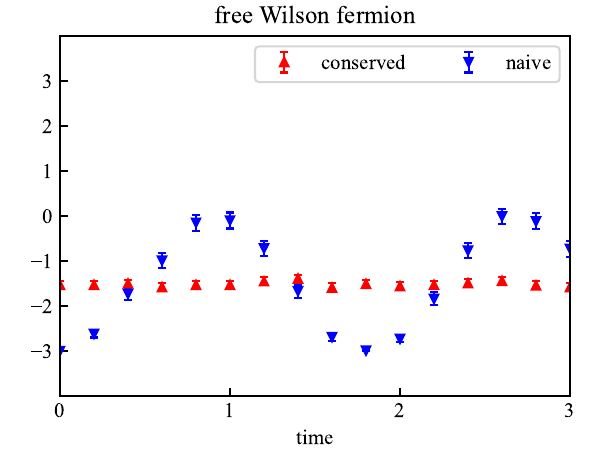}
\end{minipage}
\hfill
\begin{minipage}{0.5\linewidth}
\centering
\includegraphics[scale=0.75]{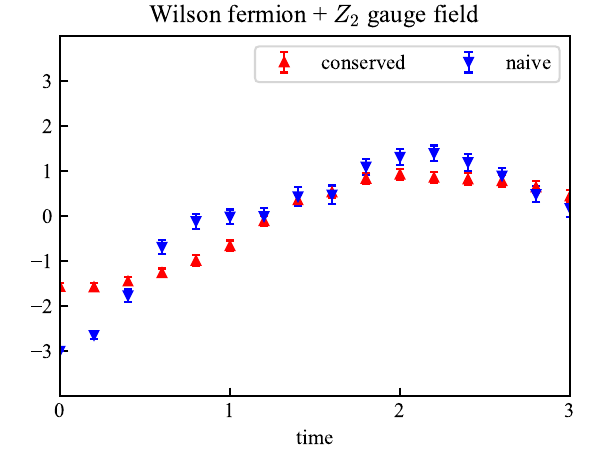}
\end{minipage}
\caption{\label{fig2}
Time evolution of the conserved and naive chiral charges \cite{Hayata:2023skf}.
}
\end{figure}

Figure \ref{fig2} shows the simulation results obtained by a noiseless emulator.
The simulation was done on a periodic three-site lattice.
The left panel is the time evolution of the free Wilson fermion.
As explained above, the naive chiral charge $\langle Q_{\rm naive} \rangle$ is not conserved and the conserved chiral charge $\langle Q \rangle$ is really conserved.
The right panel shows the interacting case.
Although we are eventually interested in continuous gauge theory, such as QED or QCD, here we consider the $Z_2$ lattice gauge theory.
We write down the conserved chiral charge operator
\begin{equation}
 Q = \sum_x \bigg\{ \frac{1}{2} \psi^\dagger(x) \gamma^5 \psi(x)
+ \frac{1}{4} \psi^\dagger(x) \gamma^5 (1-\gamma^1) \sigma_3(x) \psi(x+1) 
+ \frac{1}{4} \psi^\dagger(x+1) \gamma^5 (1+\gamma^1) \sigma_3(x) \psi(x) \bigg\},
\end{equation}
the fermion Hamiltonian
\begin{equation}
 H_f = \sum_x \bigg\{ \psi^\dagger(x) \gamma^0 \psi(x) -\frac{1}{2} \psi^\dagger(x) \gamma^0 (1-\gamma^1) \sigma_3(x) \psi(x+1)
- \frac{1}{2} \psi^\dagger(x+1) \gamma^0 (1+\gamma^1) \sigma_3(x) \psi(x) \bigg\} ,
\end{equation}
and the gauge field Hamiltonian
\begin{equation}
 \hat{H}_g = - \sum_x \sigma_1(x).
\end{equation}
The Pauli matrices, $\sigma_3(x)$ and $\sigma_1(x)$, are gauge field operators.
The chiral charge operator commutes with the fermion Hamiltonian, $[Q,H_f]=0$, but does not commute with the gauge Hamiltonian, $[Q,H_g]\neq 0$, due to the Pauli matrices.
Because of this non-commutativity, the chiral charge shows non-trivial time evolution, as seen in the figure. 
The further interpretation of this time evolution is unclear in the $Z_2$ lattice gauge theory because the theory does not have the continuum limit.
Nevertheless, our demonstration is successful.
The conserved chiral charge works well.
It is conserved for the free fermion and dynamically generated by the gauge interaction.
When we apply this analysis to the quantum simulation of one-dimensional QED or QCD, we will be able to study dynamical chirality generation by the chiral anomaly, probably, in the near future.
In Fig.~\ref{fig2}, the naive chiral charge shows a similar behavior, but we should not use it because we cannot distinguish the physical effect and the lattice artifact.

\section*{Acknowledgments}

This work was supported by JSPS KAKENHI Grant No.~19K03841, 21H01007, and 21H01084.

\end{document}